# Amplitude- and phase-resolved nano-spectral imaging of phonon polaritons in hexagonal boron nitride

*Zhiwen Shi[1]‡, Hans A. Bechtel[2]‡, Samuel Berweger[3,4], Yinghui Sun[1], Bo Zeng[1], Chenhao Jin[1], Henry Chang[1], Michael C. Martin[2], Markus B. Raschke[4], Feng Wang[1, 5, 6]\**

[1]Department of Physics, University of California at Berkeley, Berkeley, California 94720, USA.

[2]Advanced Light Source Division, Lawrence Berkeley National Laboratory, Berkeley, California 94720, USA.

[3]National Institute of Standards and Technology, Boulder, CO 80305

[4]Department of Physics, Department of Chemistry, and JILA, University of Colorado, Boulder, Colorado 80309, USA.

[5]Materials Sciences Division, Lawrence Berkeley National Laboratory, Berkeley, California 94720, USA.

[6]Kavli Energy NanoSciences Institute at the University of California, Berkeley and the Lawrence Berkeley National Laboratory, Berkeley, California, 94720, USA.

ABSTRACT:

Phonon polaritons are quasiparticles resulting from strong coupling of photons with optical phonons. Excitation and control of these quasiparticles in 2D materials offer the opportunity to confine and transport light at the nanoscale. Here, we image the phonon polariton (PhP) spectral response in thin hexagonal boron nitride (hBN) crystals as a representative 2D material using amplitude- and phase-resolved near-field interferometry with broadband mid-IR synchrotron radiation. The large spectral bandwidth enables the simultaneous measurement of both out-of-plane (780 cm$^{-1}$) and in-plane (1370 cm$^{-1}$) hBN phonon modes. In contrast to the strong and dispersive in-plane mode, the out-of-plane mode PhP response is weak. Measurements of the PhP wavelength reveal a proportional dependence on sample thickness for thin hBN flakes, which can be understood by a general model describing two-dimensional polariton excitation in ultrathin materials.

TEXT:

Phonon polaritons (PhPs) result from coupling of photons with optical phonons in polar crystals. Unlike *plasmon* polaritons[1-3] that usually span a very broad energy range, PhPs provide a spectrally selective response related to the optical phonon modes from infrared to THz spectral range. The PhPs can have strong spatial confinement, and may enable potential applications for enhanced IR light-matter interaction[4, 5], high density IR data storage[6], coherent thermal emission[7], development of metamaterials[8, 9] and frequency-tunable terahertz wave generation[10].

Hexagonal boron nitride (hBN), is a convenient model system for studying PhPs in ultrathin materials, because (1) hBN flake with different thickness is easy to achieve due to its 2D nature; (2) hBN flakes are of extremely high quality with no dangling bonds. It has two phonon modes in the mid-infrared region: the low frequency out-of-plane (~780 cm$^{-1}$) mode and the high-frequency in-plane (~1370 cm$^{-1}$) mode, both of which have negative real parts of the dielectric function and thus have the potential to support PhPs[11]. Furthermore, hBN is a promising material for nanotechnology applications due to its two-dimensional (2D) layered structure, excellent electrical insulation, and chemical and thermal stability. It has attracted great interest as a substrate for high mobility graphene[12-16], an ideal dielectric layer and spacer for 2D heterostructures[17, 18], as well as for its intrinsic UV lasing response[19].

Recently, several groups studied hBN PhP related behaviors[20-23]. Especially, Dai *et al.* studied PhPs in thin hBN flakes with laser-based scattering scanning near-field optical microscopy (*s*-SNOM), and observed a thickness-dependent PhP dispersion[20]. However, that study focused on the upper spectral range around the hBN in-plane (~1370 cm$^{-1}$) vibration phonon mode with only amplitude information of the coherent PhP response, and relied on complex modeling to explain the thickness-dependent PhP dispersion. Here, we use broadband synchrotron infrared nanospectroscopy (SINS)[24] for amplitude and phase-resolved spatio-spectral imaging of the PhPs in thin hBN flakes covering both the low-frequency out-of-plane (~780 cm$^{-1}$) and high-frequency in-plane (~1370 cm$^{-1}$) characteristic phonon modes of hBN. We observe a strong PhP response only for the in-plane phonon mode. We describe the observed standing wave PhP behavior using a simple cavity model to simultaneously fit the amplitude and phase[25], which allows us to extract the wavelength and reflection coefficients as they relate to the dispersion relation. The measured phonon polariton wavelength is proportional to the sample thickness as described by a simple theory general to 2D polaritons in ultrathin layered materials.

Near-field Fourier transform infrared nano-spectroscopy and imaging was performed with SINS at the Advanced Light Source[24]. In brief, broadband synchrotron IR light is focused onto the apex of a metal-coated atomic force microscope (AFM) tip with curvature radius $r \approx 30$ nm (Fig. 1A). The resulting large near-field momentum provided by the spatial field localization at the apex extends into the typical momentum range of hBN phonon polaritons and allows for their optical excitation. The excited polariton wave propagates radially outward from the near-apex launched region along the hBN surface (Fig. 1B). When reaching the sample edge, the polariton waves are reflected. The back-reflected polariton waves interfere with the excitation field at the tip and modify the amplitude and the phase of the tip-scattered near-field, which is detected by interferometric heterodyne amplification[24]. This allows for the spatial and spectral imaging of the PhP waves. The strong dependence of the PhP wavelength on excitation frequency leads to a variation in the observed spectrum, as shown in Fig. 1C with three representative amplitude spectra taken at different distances from the edge of a d=75 nm thick hBN sample.

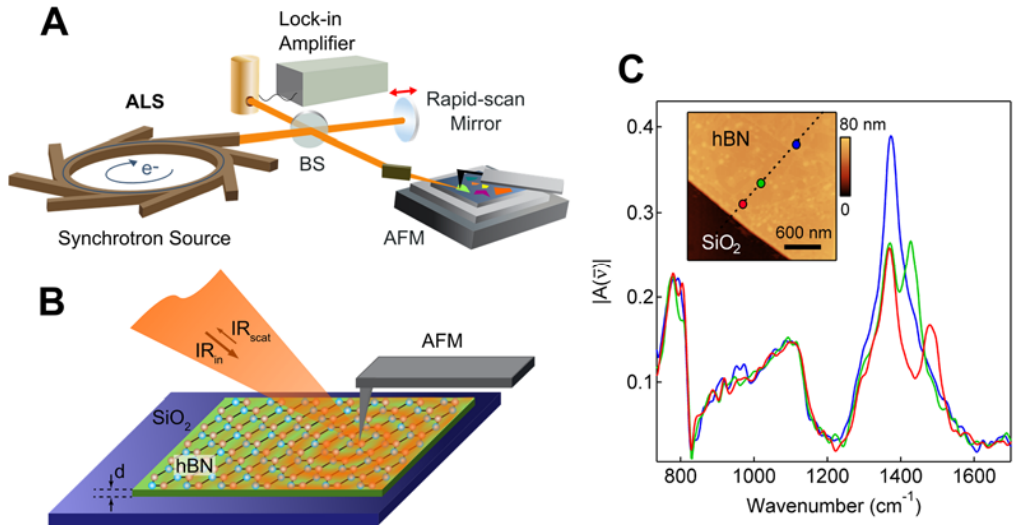

**Figure 1.** (A) Schematic of synchrotron infrared nano-spectroscopy (SINS) with the Advanced Light Source (ALS) as the synchrotron source, a KBr beamsplitter, and an atomic force microscope (AFM). (B) Broadband synchrotron infrared light illuminates the conductive AFM tip, which launches polariton waves that propagate radially outward from the tip along the hBN surface and are reflected off the edges. (C) Representative amplitude $|A(\bar{\nu})|$ spectra taken from three different positions (marked with colored dots) of a 75 nm thick hBN flake, shown in the inset. A peak appears on the high-frequency side of the main phonon peak in the spectra taken near the hBN edge, due to the interference of the forward moving and backward moving polariton wave reflected by the hBN edge.

To prepare samples, hBN[26] is directly exfoliated onto a $SiO_2$/Si (300nm/500μm) substrate with adhesive tape. The achieved hBN flake thickness varies from several atomic layers to hundreds of nanometers. After taking an AFM image of an hBN flake to determine its thickness, we perform either a single SINS line scan perpendicular to a straight edge or record full spatio-spectral area scans in the edge region. Several specimens with thickness varying from 3 nm to 160 nm were measured and analyzed.

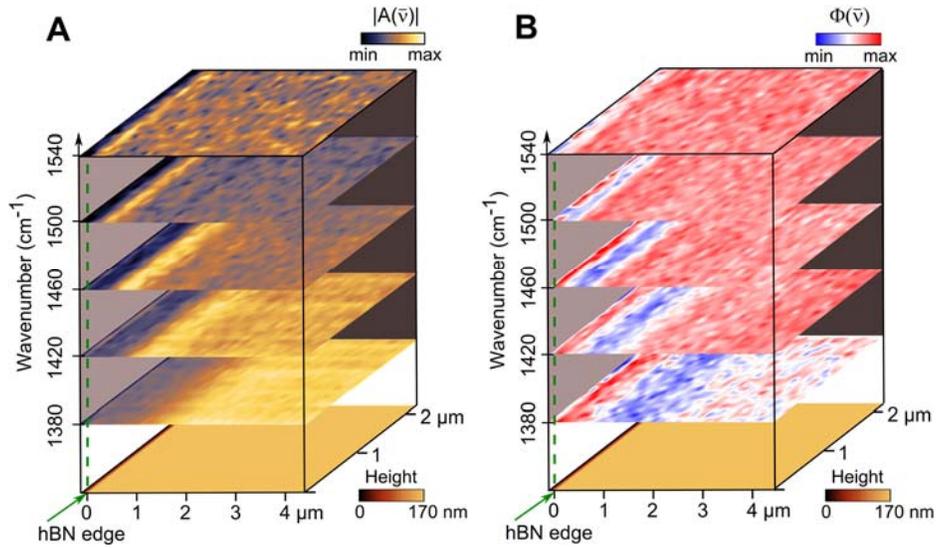

**Figure 2.** Sequence of real-space SINS images of near-field (A) amplitude $|A(\bar{v})|$ and (B) phase $\Phi(\bar{v})$ near an edge of a 160 nm-thick hBN flake acquired simultaneously. The bottom image in both (A) and (B) is the topography of the sample, showing the sample edge on the left side of the image stack at x = 0 (indicated by the dotted green line). The color map for the amplitude and phase images have been individually normalized to the maximum and minimum amplitude and phase value of each image for better visibility. Interference between the forward-propagating and back-reflected polariton waves modulates the near-field amplitude and phase as a function of the distance from the hBN edge. Different interference patterns are due to different polariton wavelengths.

Figure 2A and 2B shows the real-space two-dimensional (2D) mapping of the spectral near-field amplitude $|A(\bar{v})|$ and phase $\Phi(\bar{v})$ of a 160 nm-thick hBN flake for selected excitation frequencies, respectively. The bottom images of Fig. 2A and 2B are the topography of the sample; the straight edge is located on the left side (x = 0 μm). Interference between the forward-propagating and back-reflected polariton waves modulates the near-field amplitude and phase as a function of the distance from the hBN edge. Different interference patterns reflect different polariton wavelengths. All the interference patterns are parallel to the sample edge, showing uniform reflection of the PhP wave at the edge. Progressing from the top image to the bottom image, the maximum amplitude (Fig. 2A) and minimum phase (Fig. 2B) gradually move away from the hBN edge, due to the increase in polariton wavelength increases with decreasing incident light frequency.

The large bandwidth of the synchrotron light enables us to measure the PhP IR response in a range that covers both in-plane and out-of-plane phonon modes. Figure 3A and 3B shows broadband spatio-spectral maps of near-field amplitude and phase, respectively, for a line scan perpendicular to the edge of the sample. Increased scattering amplitude (Fig. 3A) around 1370 cm$^{-1}$ and 780 cm$^{-1}$ correspond to the in-plane and out-of-plane phonon modes in hBN, and the spectral feature around 1100 cm$^{-1}$ is due to the SiO$_2$ phonon of the substrate. Spatially dependent features arising from propagating polaritons only show up around the in-plane phonon mode between 1370 cm$^{-1}$ and 1610 cm$^{-1}$ (Fig. 3A and 3B) (The reason why the out-of-plane mode PhPs were not observed will be discussed below). This phenomenon is revealed more clearly in Fig. 3C and 3D with five sections corresponding to spectra at the positions marked with dashed lines in Fig 3A and 3B, respectively. Peaks in Fig. 3C at 780 cm$^{-1}$, 1100 cm$^{-1}$ and 1370 cm$^{-1}$ remain unchanged, while the interference peak appears on the right shoulder of the main phonon peak at 1370 cm$^{-1}$ and progressively moves to higher frequency when the tip approaches the sample edge. Similar behavior is also observed for the near-field phase (Fig. 3D).

In the following, we describe the observed near-field amplitude and phase self-consistently using a simple cavity model. As discussed in detail elsewhere[25], we model the complex-valued substrate response $\widetilde{\Psi}$ from the dielectric sample response $\widetilde{\psi}_{hBN}$, the resonant local PhP response $\widetilde{\psi}_{PhP,0}$, and the reflected PhP $\widetilde{\psi}_{PhP,1}$ as $\widetilde{\Psi} = \widetilde{\psi}_{hBN} + \widetilde{\psi}_{PhP,0} + \widetilde{\psi}_{PhP,1}$. To account for the spatial averaging by the tip we convolve the sample response with a weighting function Θ to yield the modeled signal $\tilde{A}(r) = \tilde{\Psi}(r') * \Theta(r - r')$. To approximate the tip geometry we use a Gaussian with a width of 30 nm for $\Theta$.

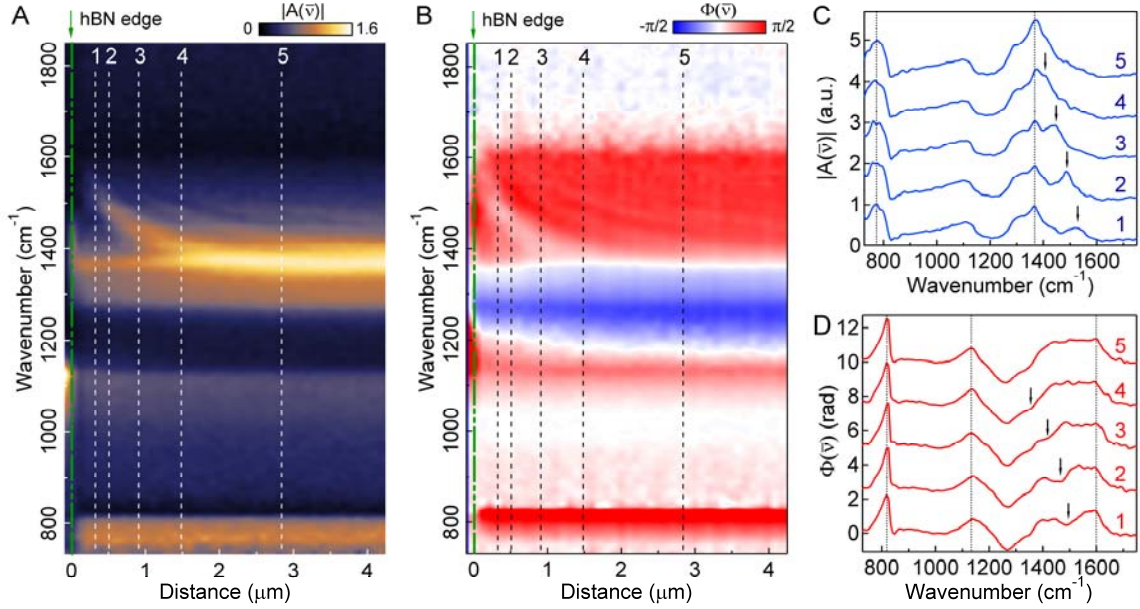

**Figure 3.** Spatio-spectral SINS line scan of hBN phonon polaritons with spectral cuts. Simultaneously acquired (A) near-field amplitude $|A(\bar{\nu})|$ and (B) phase $\Phi(\bar{\nu})$ linescan obtained perpendicular to the edge of a 160 nm-thick hBN flake. The green dashed line at x = 0 marks the edge of the hBN crystal. Signals around 1370 cm$^{-1}$ and 780 cm$^{-1}$ correspond to in-plane and out-of-plane phonon mode in hBN, respectively, and the signal around 1100 cm$^{-1}$ is related to the SiO$_2$ phonon mode of the substrate. Strong PhP features only appear around the in-plane phonon mode (1370 cm$^{-1}$). Near-field (C) amplitude and (D) phase spectra at five representative positions marked in (A) and (B).

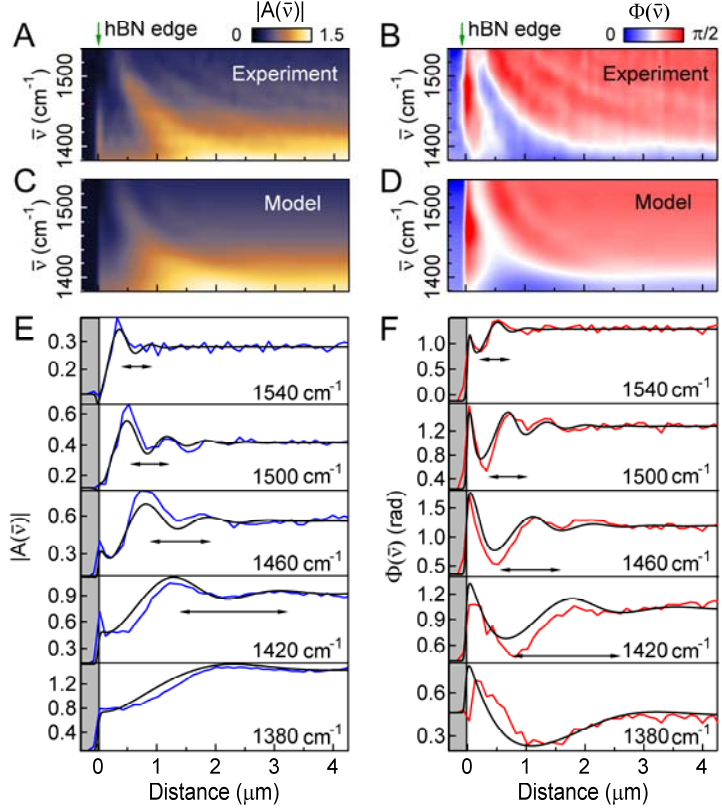

**Figure 4.** Comparison between experiment and cavity model with spatial profiles. Experimental (A) SINS amplitude $|A(\bar{\nu})|$ and (B) phase $\Phi(\bar{\nu})$ spatio-spectral linescan in the region of hBN in-plane phonon mode of a 160 nm thick hBN flake. Note the different color scales from Fig. 3 with absolute phase values. Simulated SINS (C) amplitude and (D) phase spatio-spectral linescan of the same hBN flake in (A) and (B) using the cavity model discussed in the text. The color scales for the experimental and simulated linescans are identical. (E) Experimental (blue) and simulated SINS amplitude (black) distance profiles extracted from (A) and (C), respectively. (F) Experimental (red) and simulated SINS phase (black) distance profiles extracted from (B) and (D), respectively.

We then fit the wavelength-dependent amplitude and phase simultaneously using this cavity model. Shown in Fig. 4A and 4B are the experimental PhP response from 1380 - 1540 cm$^{-1}$. Fig. 4C and 4D show the corresponding results of the fit applying the cavity model. Further shown in Fig. 4E and 4F are line cuts at select frequencies with the experimentally measured distance dependence in blue and red, respectively, and the modeled data (black). From the fitting, we can estimate the polariton wavelength, reflection coefficients, and damping. We typically find a reflection coefficient of $R = -1$ (corresponding to a reflection phase of $\pi$) at the hBN edge. However, for the case of lower frequency PhP modes the increase in amplitude near the edge as seen in Fig. 4E suggests the possibility of a change in reflection coefficient. In addition to the tip, the edge itself might contribute to the generation of PhP. In contrast to the much stronger polarizability of the metallic tip, this contribution is weak and neglected in our analysis. However, further theoretical investigation will be required for a complete description of the PhP response and to fully understand the PhP reflection and its phase behavior at the hBN edges across the whole spectral range.

Besides the 160nm thick sample, we systematically measured a set of samples with thicknesses from 3 nm to 160 nm. Spatio-spectral maps of near-field amplitude and phase for three more samples with thickness of 7 nm, 22 nm and 40 nm are displayed in Fig. 5A and 5B. The gradual change of the PhP feature with sample thickness is clearly shown and is caused by the variation of polariton wavelength with BN crystal thickness. Figure 5C shows the dependence of the polariton wavelengths $\lambda_p$ on sample thickness $d$ for three selected frequencies at $\bar{\nu} = $ 1410, 1440 and 1470 cm$^{-1}$, exhibiting linear scaling behavior for all frequencies.

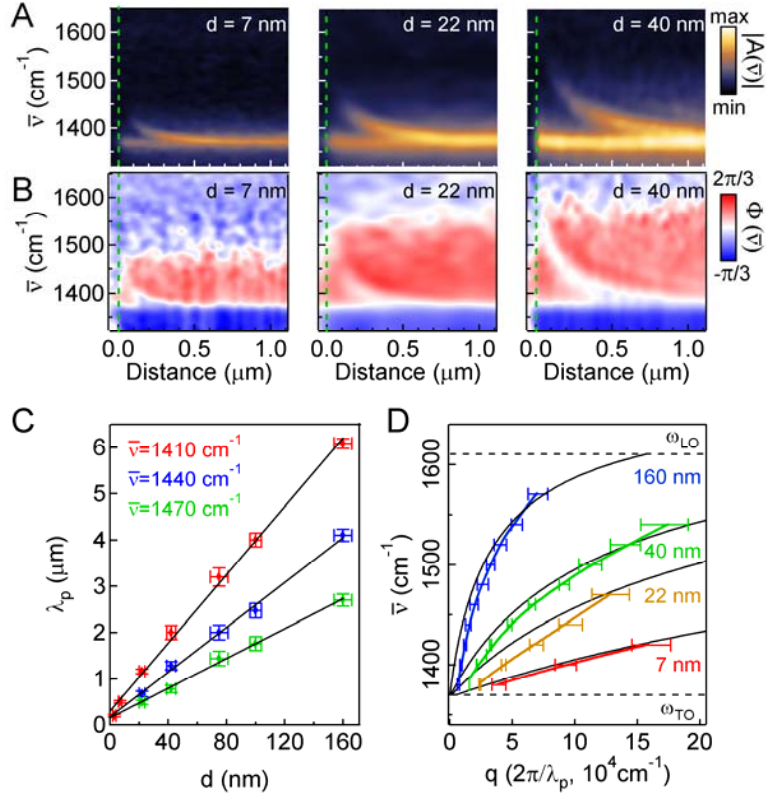

**Figure 5.** PhP dispersion and evolution of the phonon polariton wavelength with sample thickness Spatio-spectral maps of (A) SINS amplitude $|A(\bar{v})|$ and (B) phase $\Phi(\bar{v})$ for samples with thickness of 7nm, 22nm, and 40nm. (C) Evolution of the phonon polariton wavelength, $\lambda_p$, with sample thickness, d, at three representative excitation frequencies: 1410 cm$^{-1}$, 1440 cm$^{-1}$ and 1470 cm$^{-1}$. For thin hBN flakes, $\lambda_p$ is proportional to sample thickness. (D) Experimental (colored curves) and theoretical (black curves) dispersion relations of PhPs on hBN flakes with different thicknesses.hBN is a natural hyperbolic material, whose in-plane and out-of-plane real part of dielectric functions are of opposite sign in certain energy range related to its phonon modes[27]. In bulk hBN, many waveguide modes of phonon polaritons can exist and their description requires the full knowledge of both the in-plane and out-of-plane dielectric constant[20, 21]. However, in the 2D limit, where the hBN flake thickness can be a few tens of nanometers or even a few atomic layers, the only observable mode in our near-field microscopy

study is the fundamental mode. All the other modes will have wavelengths shorter or comparable to the film thickness because they require at least one full oscillation over the film thickness. The higher-order polaritons, due to their very short wavelengths, are strongly confined in the bulk and do not couple efficiently to the external probes. That is why they have not been observed in our near-field optical study.

To understand the experimentally observed scaling behavior, we present a general model describing 2D polaritons in ultrathin materials where the thickness $d$ is much smaller than the polariton wavelength $\lambda_p$. 2D polaritons are characterized by in-plane charge density oscillations, which generate an appreciable electrical field distribution near the surface that can couple efficiently to a tip via its near-field. The elementary charge oscillations can arise from free electrons, ions, and excitons, known respectively as plasmon, phonon, and exciton polaritons. In contrast, out-of-plane charge density oscillations generate large electrical field inside the 2D material, but the field outside the material is negligible (at the level of $d/\lambda_p$). Such out-of-plane modes, like the 780 cm$^{-1}$ phonon in hBN, do not couple efficiently to a near-field probe, and thus are less easily observed in our experiment.

For 2D polaritons associated with in-plane charge oscillations, the physics is uniquely determined by a single parameter—the in-plane 2D susceptibility $\chi^{2D} = \chi_{\parallel}^{bulk} \cdot d = (\varepsilon_{\parallel}^{bulk} - 1) \cdot d$, because the electrical field for the fundamental PhP mode in ultra-thin hBN flakes is mainly in plane. Here $\chi_{\parallel}^{bulk}$ and $\varepsilon_{\parallel}^{bulk}$ are the in-plane bulk susceptibility and dielectric constant, respectively. The dispersion relation of 2D polaritons (in the long-wavelength limit) is simply determined by a zero dynamical screening function $1 + \varepsilon_0 \frac{q^2}{e} \cdot \chi^{2D}(\omega) \cdot v_c^{2D}(q) = 0$ [28], where $v_c^{2D}(q) = e/(2\varepsilon_0 \varepsilon_{eff} \cdot q)$ is the 2D Coulomb interaction, q=$2\pi/\lambda_p$ is the polariton wavevector, and $\varepsilon_{eff}$ is the effective dielectric constant due to the environmental screening. Consequently, we obtain a simple yet general solution for 2D polariton dispersion in ultrathin films $\lambda_p = -\frac{\pi d}{\varepsilon_{eff}} \cdot [\varepsilon_{\parallel}^{bulk}(\omega) - 1]$. This simple model correctly predicts the linear scaling behavior between the phonon polariton wavelength $\lambda_p$ and the thickness $d$ for a given frequency in thin hBN crystals (Fig. 5C, here the value of hBN dielectric function $\varepsilon_{\parallel}^{bulk}(\omega)$ is from ref. 11). It also shows that a 2D phonon polariton is independent of the perpendicular dielectric constant $\varepsilon_{\perp}^{bulk}(\omega)$ in thin crystals. Using an effective environmental dielectric constant $\varepsilon_{eff} = (\varepsilon_{air} + \varepsilon_{SiO_2})/2$, our theory qualitatively reproduce the experimentally observed dispersion relation (Fig. 5D). The deviation between the theory and the experimental data mainly originates from the inaccuracy of $\varepsilon_{eff}$ from neglecting the silicon substrate (300nm away from hBN layer) contribution, which is polariton wavelength dependent. The theoretical analysis is true only when the sample thickness is much smaller than the polariton wavelength. In addition to PhPs in hBN, we expect this simple model to provide a general description on any 2D polaritons, including plasmon polaritons in graphene as well as phonon or exciton polaritons in other ultrathin crystals.

In summary, we have investigated phonon polaritons in hBN flakes with different thicknesses by using near-field spectral imaging with broadband synchrotron infrared light. We find that in the measured hBN basal plane, a strong PhP response is only observed around the in-plane phonon mode and the measured polariton wavelength is proportional to sample thickness. The scaling behavior can be well understood with a simple theory that is universal for polaritons in thin 2D materials. Our results open up new avenues in engineering infrared light at the nanometer scale for novel photonic nanodevices and understanding intriguing polariton behaviors in low dimensional nanomaterials.

**Methods:**

Broadband synchrotron IR light from the Advanced Light Source (ALS) at Lawrence Berkeley Laboratory is focused onto the apex of an oscillating platinum silicide (Nanosensors, PTSi-NCH)[#] tip of a modified AFM (Innova, Bruker)[#], and the scattered light is detected by an MCT detector. The near-field signal is discriminated from the far-field background by lock-in detection at twice the tip-oscillation frequency[29, 30]. Spectra are acquired with interferometric heterodyne detection using a modified commercial FTIR spectrometer (Nicolet 6700, Thermo-Scientific)[#] configured as an asymmetric Michelson interferometer[24]. The broad spectral range of the synchrotron source enables the acquisition of spectra across the entire mid-IR range (700-5000 $cm^{-1}$) within a single scan. The Fourier transform of the resulting interferogram results in the complex-valued near-field spectrum. The real and imaginary spectra, typically represented as spectral amplitude and phase, relate to the complex dielectric function of the material[30, 31].

AUTHOR INFORMATION

**Corresponding Author**

* Email: fengwang76@berkeley.edu (F.W.)

**Author Contributions**

‡Z. S. and H. A. B. contributed equally to this work.

**Notes**

The authors declare no financial competing interests. [#] Mention of commercial products is for informational purposes only, it does not imply NIST's recommendation or endorsement.

ACKNOWLEDGMENT

Sample preparation and optical measurements in this work were mainly supported by the Office of Naval Research (award N00014-13-1-0464). F.W. acknowledges support from a David and Lucile Packard fellowship. The ALS is supported by the Director, Office of Science, Office of Basic Energy Sciences, and the BSISB is supported by the Office of Biological and Environmental Research, all through the U. S. Department of Energy (DOE) under Contract No. DE-AC02-05CH11231. M.R. acknowledges supported by the U. S. DOE, Office of Basic Energy Sciences, Division of Materials Sciences and Engineering under Award No DE-FG02-12ER46893.

ABBREVIATIONS

hBN, hexagonal boron nitride; PhP, phonon polariton; SINS, synchrotron infrared nanospectroscopy; AFM, atomic force microscope.